\documentclass[a4paper,fleqn,usenatbib]{mnras}
\usepackage{amssymb,amsmath,graphicx,psfrag,mathtools}
\usepackage[T1]{fontenc} 
\usepackage{ae,aecompl} 
\usepackage{subfigure}
\usepackage{url}
\title[Circumsolar Rings Occulting Boyajian's Star?]{Can Dips of Boyajian's Star be Explained by Circumsolar Rings?}
\author[J. I. Katz]{
J. I. Katz,$^{1}$\thanks{E-mail katz@wuphys.wustl.edu} 
\\
$^{1}$Department of Physics and McDonnell Center for the Space Sciences,
Washington University, St. Louis, Mo. 63130 USA 
}
\date{Accepted XXX.  Received YYY; in original form ZZZ} 
\pubyear{2017} 
\date{\today}
\begin{document} 
\label{firstpage} 
\pagerange{\pageref{firstpage}--\pageref{lastpage}} 
\maketitle 
\begin{abstract}
Could the dips of ``Boyajian's Star'' (KIC 8462852) have been caused by
matter in our Solar System?  The interval between periods of deep dips is
nearly twice the orbital period of the Kepler satellite.  I consider
heliocentric obscuring rings in the outer Solar System that graze
the line of sight to the star once per Kepler orbit.  The hypothesis
predicts that future dips may be observed from Earth during windows
separated by a year, although their detailed structure depends on the
distribution of particles along the ring.  Dips observed at terrestrial
sites separated by $> 1000$ km in a direction perpendicular to the Earth's
orbital motion may be uncorrelated.
\end{abstract}
\begin{keywords} 
Kuiper belt: general, stars: peculiar, (stars:) planetary systems 
\end{keywords} 
\setcounter{footnote}{0}
\section{Introduction}
``Boyajian's star'' (KIC 8462852) is a peculiar variable discovered by the
Kepler exoplanet-finding mission \citep{B10,K10}.  { \citet{B16,WS16}
review these and other observations and contain extensive references to the
literature.}  Spectroscopically an ordinary F3V dwarf, Boyajian's star
shows remarkable deep dips.  These are much too deep (up to 21\%) to have
been caused by planetary transits, for the radius of the transiting objects
would have to be nearly half ($\sqrt{0.21} \times $) the radius of the F3V
star, or at least $5 \times 10^{10}$ cm. This is several times larger than
the radius of any cold object, or even than the radii of ``hot jupiters''.
Planetary transits are also excluded because the dips are not periodic.

This behavior has been interpreted \citep{B16} as the result of obscuration
near Boyajian's star by dust clumps bound to multiple planetary
objects or the transient result of breakup of comet-like objects.  The
clumps would not only have to be opaque but to have at least half the
diameter of the star.  This explanation is constrained by the absence of an
infrared excess that would be produced by warm dust, and leaves many
questions unanswered \citep{B16}.  Other models that place the obscuring
matter close to the star, such as a nearly edge-on disc with eruptions that
interpose absorbing or scattering matter in our line of sight, are also
difficult to reconcile with all the data and with our understanding of their
dynamics (a disc eruption would { put} disc matter in an inclined orbit, and
would lead to periodic obscuration, mimicking a planet).

A single 16\% dip several days long, but with half width at half minimum
(HWHM) about 0.5 day, is separated from a complex cluster of dips, including
dips of 3\%, 8\% and 21\%, by about 750 days.  This interval is close to
twice the $P_K = 372.53$ d orbital period of the Kepler satellite
\citep{B10,K10,WS16}, which hints at a Solar System origin.

The dips qualitatively resemble the occultations of stars by planetary rings
that led to the discovery of the rings of Uranus \citep{EDM77} and then of
all the major planets.  { Here I consider the hypothesis that the matter
obscuring Boyajian's star is not orbiting that star, but rather is
found in the Solar System, so that the motion of the observer introduces an
annual parallax and periodicity \citep{WS16}.  I suggest that the obscuring
matter orbits on heliocentric rings.  The distribution of matter along these
rings is likely patchy rather than continuous, observationally constrained
by the occurrences, shapes and widths of the dips.  If this is correct, then
the dips of Boyajian's star are exemplars of the occultation astronomy
suggested by \citet{D92}.

In this picture Boyajian's star is unique because its ecliptic latitude
aligns it with matter in the remote outer Solar System, and specifically with
the direction of tangency to orbits of this distant matter.  Tangency
increases the likelihood and prolongs the duration of deep absorption.}
The observed deep dips have durations of several hours that cannot be
explained by non-tangent passages through narrow rings, while broad rings
are excluded by the fact that only one of $\approx$ 150,000 Kepler stars
shows such dips.
\section{The Dips}
\begin{figure}
\centering
\includegraphics[width=3.3in]{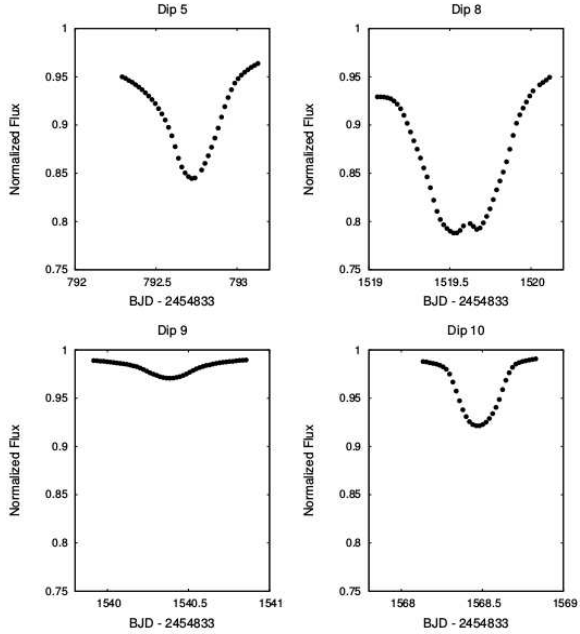}
\caption{\label{dips} The four deep ($> 1\%$) { numbered} dips of
\citet{B16}, all plotted on the same scale to facilitate comparison.  These
dips have comparable, but not equal, widths, and dip 8 can be described as
two overlapping narrower dips.  Single data are missing near the minima of
dips 5 and 8 and near day 1520.0 in dip 8
\citep{B10,K10}.\protect\footnotemark}
\end{figure}

Fig.~\ref{dips} shows the deep dips in greater detail than in \citet{B16}.
The dips are smooth on the data sampling time scale of 0.02 d and have
comparable, but not identical, widths.  { The deepest} dip (8), with a
double minimum, { may be the closest to exact tangency (see
Sec.~\ref{doubledip} for a discussion)}.

\footnotetext[1]{\url{https://exoplanetarchive.ipac.caltech.edu/applications/ETSS/Kepler\_index.html} accessed May 16, 2017.}

{ If the geometry were known exactly, the radial distribution of
absorption within a ring defined by the particles' orbits could be
deconvolved from the shapes of the dips.  The fact that their shapes, HWHM,
depths and the comparative strengths of their shoulders and penumbr\ae\
differ indicate that tangency is (unsurprisingly) not exact, precluding such
deconvolution.  The properties of a ring may also vary along its length, 
like the rings filled by disrupted cometary material that produce meteor
showers.}

\begin{figure}
\centering
\includegraphics[width=3.3in]{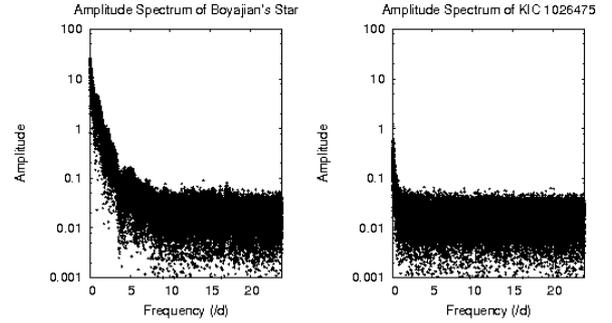}
\caption{\label{FFT} The Fourier transforms of the raw data records of
Boyajian's Star and KIC 1026475 from BJD-2454833 = 251.805--1591.001 are
shown up to the Nyquist frequency.  KIC 1026475 resembles Boyajian's star in
magnitude and $T_{eff}$ and shows the intrinsic noise in the Kepler data.
The first 131 days of the records are excluded in order to have exactly
$2^{16} = 65536$ evenly spaced measurement intervals for the Fast Fourier
Transform.  The 5885 missing data are linearly interpolated.  The rotation
frequency of Boyajian's star and its second and third harmonics are visible
as shoulders at 1.137/d, 2.274/d and 3.410/d, respectively
\protect\citep{B16}.  The bump around 5/d reflects the deep dips with HWHM
$\sim 0.25$ d.  The fluctuation spectrum above that frequency is almost
entirely instrumental.}
\end{figure}

\citet{B16} Fourier transformed the data with the dips excised, and in the
lower frequency part of the spectrum discovered a 0.8797 d periodicity
attributed to the rotation of the star.  A Fourier transform of the raw data
is shown in Fig.~\ref{FFT}.  Bumps or shoulders at low frequencies reflect
the rotational frequency and harmonics found by \citet{B16}.  The bump
around a frequency of 5/d reflects the HWHM of $\sim 0.25$ d of the dips
shown in Fig.~\ref{dips} (inspection of the data shows that the other
numbered dips have comparable widths).  There is little or no evidence for
narrower dips or other higher frequency temporal structure, up to the
Nyquist frequency of 24/d.  Comparison to the Fourier transform of the
brightness of KIC 1026475, a star very similar to Boyajian's except for its
temporal stability, shows essentially identical amplitudes above a frequency
of 7/d except for the slight bumps around 9/d, 12/d and 17/d visible in
Fig.~\ref{FFT}; the remainder of the fluctuation amplitude at frequencies $>
7/$d is intrinsic to Kepler.
\section{A Ring} 
\label{ring}
Boyajian's star is at a distance of about 454 pc \citep{B16}.  In the outer
Solar System at a distance (from us) of $A$ AU the cone of rays from
the stellar disc received by the Kepler telescope (or a terrestrial
telescope) has a radius of only $5 \times 10^{-5} A_{5000}$ of the radius of
the star, or about $60 A_{5000}$ km, where $A_{5000} \equiv A/5000$.  An
obscuring object that size would be sufficient { geometrically} to block all
the light, but briefly, {although for smaller values of $A$ (such as $A
\sim 50$ in the Kuiper belt) diffraction would be important and the blockage
would be lengthened and shallower \citep{D92}.}

The orbital motion of Kepler (or the Earth) requires an obscuring cloud to
extend along the ring a distance equal to to the distance the telescope
travels (with a small correction for the motion of the obscuring matter)
during the width of the deep part of a dip $\delta t \sim 0.25$ d, or
$v_{orb} \delta t \sim 6 \times 10^{10}$ cm, where $v_{orb} \approx 3 \times
10^6$ cm/s is the difference between the observer's and the ring's orbital
velocities.  Penumbr\ae\ of $\sim 10^{12}$ cm are required to explain the
shallower broad dips.  Rather than a single object, there must be a cloud of
co-moving fragments, extended along the orbit.

\begin{figure}
\centering
\includegraphics[width=3in]{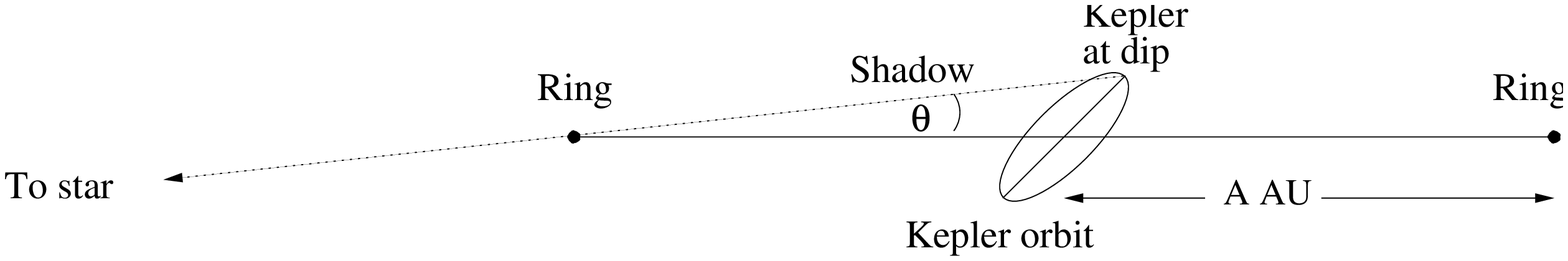}
\caption{\label{ringgeom} Nominal geometry of Boyajian's Ring, not to scale.
The angle $\theta \approx (1/A)\sin{\mathrm{\beta_{star}}}$, where $A$ is
the ring's semi-major axis (taken as the radius of a circular orbit) in
units of the radius of the observing platform's orbit (AU for Earth, 1.0132
AU for Kepler) {and $\beta$ denotes Kepler-ecliptic latitude}.  The ring's
cross-section is {smaller than can be shown to scale}, neither its
cross-section nor its orbit need be circular, and the distribution of debris
must be patchy.  In the figure the shadow is aligned with the upper extreme
of the {observer's} orbit; analogous results would be obtained if the
ring had a slightly higher ecliptic latitude so that its shadow would be
aligned with the lower extreme of the observer's orbit.}
\end{figure}

The geometry is shown in Fig.~\ref{ringgeom}.  The ring is unevenly filled
with debris, so not all lines of sight through it are occulted.  In
principle, any star with ecliptic $\beta_{ring} - \theta \le \beta_{star}
\le \beta_{ring} + \theta$ could be occulted at some time in the observer's
(Kepler's) orbit by matter in the ring, but unless the star is very close to
the bounds of this inequality the passage through the ring's shadow will be
rapid and the probability of {detection} small.

\begin{figure}
\centering
\includegraphics[width=3.3in]{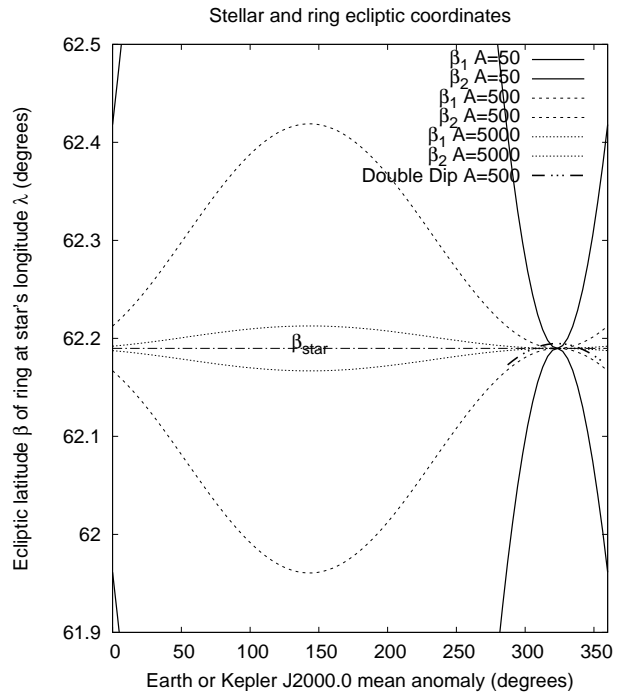}
\caption{\label{DEC} The { ecliptic latitude $\beta$} of the ring
at the right ascension of Boyajian's star as a function of { the
observer's} orbital phase, assuming a circular ring and a circular orbit of
the observer.  { These vary sinusoidally because of parallax.  The two
possible paths of a ring for each value of $A$} that maximize the time the
ring is on the line of sight between observer and star are shown.  { A
solution for $A = 500$ that gives a double dip is also shown, with its small
displacement from tangency exaggerated for clarity.}}
\end{figure}

Fig.~\ref{DEC} shows the dependence of the declination of the ring at
the right ascension of the star (in Kepler-ecliptic coordinates or in
ecliptic coordinates for an Earth-based observer) as a function of the {
observer's} mean anomaly, proportional to time for a circular ring and
circular orbit of the observer.  The { J2000.0} ecliptic coordinates of
Boyajian's star are $\lambda = 323.2111^\circ$ and $\beta =
62.1898^\circ$.  The sinusoidal curves show the two possible { ring
paths} (for each value of $A$) that maximize the time the line of sight to
the star passes through the ring, and hence maximizes the probability of
dips.  An additional curve shows how a double dip
(Sec.~\ref{doubledip} may be produced by paths that are not exactlyi
tangential.  Paths further from tangency produce shorter dips, generally
limited to a single Kepler datum, as estimated in the Discussion.

\section{Ring Parameters}
\label{parameters}
Several different estimates of the thickness of the ring can be made.  These
might indicate that there are several subrings.

If an obscuring event has HWHM $\delta t$, assumed to be centered around the
tangency of the lines in Fig.~\ref{DEC}, then the thickness of the ring
perpendicular to its orbital plane is
\begin{equation}
\label{deltaz}
\delta z \approx {1 \over 2} \left({\delta t \over P/2\pi}\right)^2 a
\sin{\theta} \approx {1 \over 2} \left({\delta t \over P/2\pi}
\right)^2 \sin{\beta_{star}}\ \mathrm{AU},
\end{equation}
where $a$ is the radius of the ring's orbit (assumed circular),
$\beta_{star}$ is the star's Kepler-ecliptic latitude and $P \approx 1$ y is
the orbital period of the platform (Earth or Kepler) from which observations
are made.  Because the ring's thickness determines the duration
$\delta t$ it shadows the observer, $\delta t$ is independent of $A$.

The smallest estimate of $\delta z$ is obtained by taking $\delta t \sim
0.25$ d, the HWHM of the narrow dips.  Then $\delta z \sim 10^8$ cm.  This
is much less than the extension of the obscuring matter along its orbital
direction, estimated (Sec.~\ref{ring}) as $\gtrsim 6 \times 10^{10}$ cm and
proportional to only the first power of $\delta t$.  That inequality is not
surprising: longitudinal dispersion in velocity spreads a cloud of debris
along its orbit to a length that increases until the orbit is filled, but
transverse velocity dispersion only produces an oscillation with amplitude
$\delta v P/2\pi$.

Larger estimates are possible.  The interval between the two deepest dips (5
and 8 in \citet{B16}) is 726.86 d, corresponding to two orbits of period of
363.43 d (the absence of a dip intermediate between these two may be
explained if the ring is patchy; see discussion following
Eq.~\ref{Tspread}).  The discrepancy between this value and Kepler's orbital
period, aside from requiring a patchy ring, indicates $\delta t \sim 5$ d.
Then $\delta z \sim 5 \times 10^{10}$ cm.  If the three deep dips 8--10 are
treated as a single event with $\delta t \sim 20$ d then the dilute
envelope of weak absorption surrounding them, and the bundle of deep dips,
are described by $\delta z \sim 10^{12}$ cm.

The multiple distinct but closely spaced dips 8--10 and the deviation of the
deep dips from accurate periodicity at Kepler's period indicate multiple
rings with nearly identical inclinations to the Kepler ecliptic but slightly
differing longitudes of their ascending nodes.
\section{Double Dip}
\label{doubledip}
{The apparent double minimum of dip 8 is suggestive of a ring whose positive
excursion in $\beta$ is to a value slightly greater (or whose negative
excursion is to a value slightly less) than that of Boyajian's star, so that
after passing through the core of the ring the line of sight to the star
passes through a lower column density of obscuring matter, then returns
through the core and its longer path length and greater column density,
before finally exiting the ring entirely.  This is illustrated in 
Fig.~\ref{DEC}.  It is unsurprising that the deepest dip is doubled,
implying passage (twice) of the line of sight through the central core of
the ring, although with only four deep dips this is not statistically
significant.

Analogously, shallower dips (the other numbered dips of \citet{B16}) may
result from lines of sight that approach but do not pass through the core of
a ring, and that therefore have single minima.  A single ring may produce
such variations if it is not exactly planar, a result of perturbations by
passing stars.}
\section{Ring Mass}
If the ring is uniformly filled with particles their mass may be
estimated
\begin{equation}
M \sim 2 \pi A \epsilon \kappa_{eff}^{-1} \delta z\,\mathrm{AU},
\end{equation}
where $\epsilon$ is the mean absorption averaged over the period over which
the line of sight to the star passes through the ring and $\kappa_{eff}$ is
the effective mass extinction coefficient of ring particles (small for large
particles) {in cm$^2$/g.  Estimating $\epsilon \sim 10^{-3}$ {from the
mean attenuation} over $\delta t \sim 20$ d if $\delta z \sim 10^{12}$
cm (Sec.~\ref{parameters}) and taking $\kappa_{eff} \sim 10^4$ cm$^2$/g
(appropriate if ring particles are micron-sized) yields $M \sim 3 \times
10^{22} A_{5000}$ g.  Smaller $\delta z$ would imply larger $\epsilon
\propto (\delta z)^{-1/2}$ but a smaller mass because the line of sight
would spend less time passing through the ring.  These values are consistent
with the breakup of a $\sim 500 A_{5000}^{1/3}$ km diameter object into fine
particles.  Masses larger, in proportion to the particle size, would be
required if the debris were coarser.

{ In the Kuiper belt ($A \sim 50$) breakup of a $\sim 100$ km diameter
body may be a plausible origin of a ring.  At much greater $A$ a ring formed
from primordial dust that never condensed into massive objects is more
plausible, and would naturally consist of efficiently attenuating sub-micron
particles.}
\section{Precession}
\subsection{Precession of the Ecliptic}
\label{preecl}
In the model proposed here a dip may occur when the ring is aligned to
the direction to Boyajian's star to an accuracy of $\delta z/(A\,\text{AU})
\approx 10^{-9} \delta z_8/A_{5000}$ radian, where $\delta z_8 \equiv \delta
z/10^8\,\text{cm}$.  The plane of the ecliptic (and of Kepler's ecliptic)
has a secular precesson rate $\approx 2 \times 10^{-6}$ radian/y
\citep{D1892}; periodic terms make this only a rough approximation.
Hence an individual ring precesses out of alignment in a time
\begin{equation}
t_{misalign} \sim {1 \over 2} \times 10^{-3} {\delta z_8 \over A_{5000}}
\ \text{y}.
\end{equation}
\subsection{Precession of the Ring}
It is surprising, and possibly an argument against this model, that
Boyajian's star is so far ($62^\circ$) out of the ecliptic; the ring axis
must make at least this angle with the ecliptic pole.  Could this be
explained by precession of the ring's orbit?  I consider three sources of
precessional torque:
\begin{enumerate}
\item The torque exerted by the concentration of mass towards the Galactic
Center.  This causes the ring's orbital axis to precess around the Galactic
symmetry axis, in analogy to the precession of the Moon's orbit as a result
of Solar torque, or of accretion discs in binary stars as a result of the
torque exerted by the mass-losing star \cite{K73}.  Quantitative results
depend on the distribution of mass in the Galaxy, but in general the ring
precession rate is
\begin{equation}
\label{GCprecess}
\Omega_{pre} \approx \Omega_{Gal}{\Omega_{Gal} \over \Omega_{ring}} \sim
2 \pi {A^{3/2} \over (2 \times 10^8\,\text{y})^2}.
\end{equation}
where $\Omega_{Gal}$ is the orbital angular velocity of the Solar System
about the Galactic center and $\Omega_{ring}$ that of the ring about the
Sun.  The ring will have precessed at least one radian in the age of the
Solar System if $A \gtrapprox 10^4$.  This is consistent with the upper end
of the preceding estimates of $A$.
\item A fraction of the Galactic mass is concentrated in the disc with a
density in the Solar neighborhood $\rho_{disc}$.  Within this layer, the
gravity gradient produces a precession rate
\begin{equation}
\label{discprecess}
\Omega_{pre} \approx \sqrt{G \rho_{disc}}\sqrt{\rho_{disc}(A\,\text{AU})^3
\over M_\odot} \approx 2 \times 10^{-24} A^{3/2}\ \text{s}^{-1},
\end{equation}
where $\rho_{disc} \approx 0.1 M_\odot$ pc$^{-3}$.  The ring has precessed
at least one radian in the age of the Solar System if $A \gtrapprox 2 \times
10^4$, consistent with the upper end of the preceding estimates of $A$.
\item Random encounters with passing masses make the plane of the ring
execute a random walk.  A rough calculation leads to a diffusion coefficient
\begin{equation}
\label{discdiffuse}
\begin{split}
D_\theta &\sim {G M_s \over b_{min}^2} {(A\,\text{AU})^3 n_s \over v_s} {M_s
\over M_\odot} \\ &\sim G \rho_{disc} {A\,\text{AU} \over v_s} {M_s \over
M_\odot} \sim 2 \times 10^{-24} {M_s \over M_\odot} A\ \text{s}^{-1},
\end{split}
\end{equation}
where $M_s$ is the mass of the perturbing objects, $v_s$ their velocity,
$n_s$ their number density, $\rho_s = M_s n_s$ and $b_{min}$ a lower cutoff
on the impact parameter, taken as the ring orbital radius $A$ AU (without
this cutoff the integral over $b$ diverges, making $D_\theta$ ill-defined,
and its effective value dependent on the single closest approach, therefore
on the duration of the period of exposure to passing masses) and taking
$\rho_s = 0.1 M_\odot$ pc$^{-3}$ and $v_s = 30$ km/s.  Then a 1 radian
change of orientation requires a time
\begin{equation}
t \sim {1 \over D_\theta} \sim 3 \times 10^{12} {M_\odot \over M_s} {1 \over
A_{5000}}\ \text{y}.
\end{equation}
This process is negligible unless $M_s \gg M_\odot$, as might be the case
if scattering is by clusters of stars or intermediate mass black holes. 
\end{enumerate}
Precession might explain the ring's inclination only if $A \gtrapprox 10^4$
or if a significant fraction of the mass passing through the Galactic disc
consists of objects of mass $\gtrsim 10^3 M_\odot/A_{5000}$.
\section{Statistics}
Boyajian's star is apparently unique among the $\approx$ 150,000 stars
studied by Kepler \citep{B16,WS16}.  Models that have an {\it a priori\/}
probability of $10^{-6}$ (for example, require a uniformly distributed
parameter to be within a fraction $10^{-6}$ of its range) may be excluded
only at the 85\% confidence level, so that they are acceptable by usual
statistical criteria; those with an {\it a priori\/} probability of
$10^{-7}$ may be excluded at the 98\% confidence level but not at the 99\%
level.  This differs from the familiar circumstance in which the brightest
or first-discovered member of a class is studied, in which case demanding
much higher {\it a priori\/} probability is appropriate.  For Kepler
objects, the class is the $\approx$ 150,000 main sequence stars studied.

{ The Kepler field of view was $\Delta \theta \approx 0.2$ radian in each
of $\beta$ and $\lambda$.  It contains only one star like Boyajian's star,
implying that the rings cover a fraction $\sim 10^{-5}$ of the field of
view.  Because rings cover narrow strips of nearly constant ecliptic latitude
(with the small sinusoidal oscillation shown in Fig.~\ref{DEC}), they must
subtend a range $\delta \beta \sim 10^{-5} \Delta \theta \sim 2 \times
10^{-6}$ rad.  Then their statistically inferred thickness $\delta z_{stat}
\lesssim A\,\delta \beta \text{\,AU} \sim 10^{11} A_{5000}$ cm.  This
is consistent with most of the estimates of Sec.~\ref{parameters}, but the
separations of dips 8--10 would require either that $A \sim 5 \times 10^4$
or that they each be produced by a separate narrow ring (or subring) in a
broader bundle of rings with similar, but not identical, orbits.  The width
of the bundle $\delta z_{bundle} \sim 5 \times 10^{10} \text{---} 10^{12}\
\text{cm} \gg \delta z$.  The wider shallow dips would be the result of a
much broader (but much less opaque) distribution of matter within this
bundle.

We can reject the hypothesis that the rings producing deep dips are
unrelated.  To explain dips 8--10 several very narrow rings must each be
oriented to produce deep dips in one star (Boyajian's) in the Kepler field
with individual {\it a priori\/} probability $p = {\cal O}$(1/150,000), but
none in any other of the 150,000 stars in the field.  If the orientations,
separated by $\sim \delta z_{bundle} \gtrsim \delta z_{stat}$, are
uncorrelated, the probability that $n$ rings are oriented to produce dips in
any one star but none in any other star is given by the Poisson distribution
for $n$ events with mean probability $np \ll 1$, and is negligible.  Hence
the rings producing dips must be correlated---part of a single bundle with
$\delta z_{bundle} \lesssim \delta z_{stat}$.

I suggest two possible explanations:
\begin{enumerate}
\item $A \gtrsim 5000$.  Then $\delta z_{stat} \sim A \delta \beta\,
\text{AU} \sim 10^{11} A_{5000}$ cm, so that $\delta z_{bundle} \lesssim
\delta z_{stat}$.  The rings are far beyond the Kuiper belt, and the entire
bundle is so narrowly confined in $\beta$ that they intercept the light of
only one star in the Kepler field.
\item Obscuring matter is concentrated in a narrow range of $\lambda$ 
(equivalently, true or mean anomaly).  Then, while a fraction $f_\beta
\sim 10^{11}\, \text{cm}/(A \Delta \theta\,\text{AU}) \sim 10^{-5}/A_{5000}$
of the Kepler stars have the right $\beta$ to be occulted, only a fraction
$f_\lambda = \delta \lambda/\Delta\theta$ also have the right $\lambda$; the
requirement that the Kepler field contain only ${\cal O}(1)$ star like
Boyajian's star implies $f_\beta f_\lambda \sim 10^{-5}$.
\end{enumerate}}
\section{Predictions}
If particulate rings explain the dips of Boyajian's star, some predictions
can be made:
\begin{enumerate}
\item Dips may be observed from the Earth when a ring grazes the line of
sight, as shown in Fig.~\ref{DEC}.  This will occur with a period $P_E =
365.25$ d.  Because $P_K \ne P_E$, the times of dips observed by Kepler
must be extrapolated back to its launch from Earth; the deep dip 5 occurred
728 days from launch, when Kepler was 16 days behind Earth in its 
Earth-trailing orbit.  Allowing for that, we predict that dips might be
observed from Earth around
\begin{equation}
\label{repetitions}
\mathrm{BJD} = 2455609 + 365.25 n,
\end{equation}
where $n = 1, 2, \ldots$, { although the apparent precision is not
justified by the observed scatter of dip timing}.

For Kepler data 365.25 d should be replaced by the Kepler year of
372.53 d.  Several shallower dips (1, 2, 3 and 6 of \citet{B16} and a dip on
May 19, 2017 \citep{ATEL17}) do not fit this pattern.  This may be
attributable to a complex ring structure with several less opaque sub-rings.
Shallow dip 4 of \citet{B16} is 366 d from deep dip 5, in reasonable
agreement.

The alignment of narrow rings, such as those inferred to produce deep
dips, with the star lasts less than a year (Sec.~\ref{preecl}), so that
annual recurrences are not expected.  For a ring complex producing extended
shallow absorption $\delta z \sim 10^{12}\,$cm (Sec.~\ref{parameters}) and
repetitions may continue for many years.  Such a complex may contain narrow
rings capable of producing deep dips, like dips 8--10 of \citet{B16}, that
have much smaller $\delta z$ and will not individually repeat.
\item The rays from the star to a point observer in the inner Solar System
are spread over a width and height $\ell_{SS} \approx 60 A_{5000}$ km  at
the ring, traversed by the observer's motion in $t_{SS} = \ell_{SS}/v_{orb}
\sim 2 A_{5000}$ s. The observed flux is an average over finer structure,
but ring structure on scales $\gtrsim \ell_{SS}$, if present, might be
observed as intensity fluctuations in sufficiently high cadence
observations.

Photometry will show time lags between terrestrial and deep space
(from future Kepler-like platforms in orbits trailing or in advance of the
Earth) observations equal to their separation along Earth's direction of
orbital motion divided by $v_{orb}$.

The rings have structure on scales $\delta z$ normal to the observer's
motion.  Dips observed from sites separated by distances $\gtrsim \delta z$
in that direction will be uncorrelated.  Because $\delta z$ may be $\sim
10^8$ cm, smaller than the size of the Earth (Sec.~\ref{parameters}), it may
be possible to observe this decorrelation between terrestrial sites.

In contrast, in circumstellar models of the dips observers anywhere in the
Solar System see the same variations, aside from small delays
attributable to light propagation time.
\item Kepler's orbit is inclined to the ecliptic by $i = 0.45^\circ$
\citep{B10,K10}.  As a result, a very narrow ring that produces dips at the
Kepler spacecraft would not produce dips at Earth, and {\it vice versa\/}.
The out-of-plane separation of these orbits $\delta z_{orb} = a \sin{i}
\sin{\chi}$, where $\chi$ is the angular separation of Earth and Kepler
measured from the Sun.  For dips 8--10 $\chi \approx 0.52$ rad, implying
$\delta z_{orb} \approx 6 \times 10^{10}$ cm.  If the ring is thicker than
this (as suggested in Sec.~\ref{parameters}) then the same dips could have
been observed from Earth.  Even for the maximum value of $\vert \sin{\chi}
\vert = 1$, $\delta z_{orb} \approx 1.2 \times 10^{11}$ cm, and for
plausible $\delta z > \delta z_{orb}$ the Kepler dip timing (allowing for
the differing orbital periods) may be extrapolated to dips observable from
Earth.
\end{enumerate}
\section{Discussion}
The occurrence of deep dips in two epochs separated by about two
Kepler-years is a hint that the phenomenon may be local rather than
circumstellar.  This evidence is suggestive but not statistically compelling
because the interval differs from an exact integer multiple of Kepler-years
by a few percent.  However, the difficulty of developing a persuasive
circumstellar model and the history of discovery of narrow planetary rings
by stellar occultation justify investigation of possible explanations
involving Solar System rings.  Preliminary estimates indicate this model is
not absurd.  

It is possible to constrain the time $T$ since the event that created the
particles if they consist of single clouds of debris moving on Keplerian
orbits since their formation, as would be plausible if they consist of
dust that never condensed to a massive object.  Assuming a width $\delta A
= \delta z/A$:
\begin{equation}
\delta\lambda \sim {3 \over 2}{\delta z \over A\,\text{AU}} A^{-3/2}
{2 \pi T \over \text{y}}\ \text{rad}.
\end{equation}
Inverting,
\begin{equation}
\label{Tspread}
T \sim 4 \times 10^9 {f_\lambda \over 0.01} {10^9\,\text{cm} \over \delta
z} A_{5000}^{5/2}\ \text{y}.
\end{equation}
{ After times $\gtrsim T$ the matter spreads into a continuous ring, but
until that happens it may be clumpy, as implied by the fact that some
expected dips are missing, such as one between dips 5 and 8.  For larger
values of $A$ ($A_{5000} \gtrsim 1$) $T$ may exceed the age of the Solar
System, explaining the survival of patchy structure.}

Matter may be distributed very nonuniformly along a ring, so that in any
Kepler-year the line of sight need not intersect a significant quantity of
absorbing matter during the brief period when the line of sight grazes the
ring.  This can explain the absence of deep dips between dip 5
and the cluster of dips 8--10 or a Kepler-year prior to dip 5.  There may be
more {\it potential\/} ``Boyajian's Stars'' among the stars observed by
Kepler that did not dip during its four years of observation because
they were viewed through gaps in a ring.  Eq.~\ref{repetitions} is a
necessary but not sufficient condition for the observation of dips in KIC
8462852 from a ring that produced dips 4, 5 and 8--10.

This model makes predictions as to the timing of possible (but only
possible, because the ring appears to be patchy,
future dips produced by the ring that produced dips 5 and 8--10.  The
existence of other rings, producing dips at other times, is not excluded.
There may be a finite, and perhaps short, correlation length between dips
observed from separate locations.  Testing this latter prediction would
provide a clear distinction between circumstellar and Solar System models
(independent of the specific ring model proposed here).  The observation of
decorrelation between Solar System observers would definitively prove a
Solar System origin, but to disprove it would require observations from at
least one distant ($\gtrsim 0.01$ AU) space platform as well as from Earth.

Ring particles move $2 \pi A^{-1/2}\,\text{AU} \sim A_{5000}^{-1/2} 10^{12}$
cm in a year (or Kepler year), so that each year the line of sight samples
an independent part of the ring, even on the penumbral scale $\sim 10^{12}$
cm estimated in Sec.~\ref{ring}.  Dips need not repeat every year.

{ Observations with higher cadence than Kepler's may reveal
non-tangential passages through a ring that last $\delta z/(v_{orb} \vert
\sin{\Delta \beta \vert})$, where $\Delta \beta$ is the difference between
the ecliptic latitude $\beta$ of the ring and the star.  This can be as
short as $\sim 30$ s for $\delta z \sim 10^8$ cm.  Non-tangential passages
might also appear in the Kepler database as shallow, single-datum, dips.
For stars in the Kepler field $\vert \Delta \beta \vert \le 0.2$ rad, and
durations would be at least a few minutes.  The absence of high frequency
noise (Sec.~\ref{FFT}) indicates that such effects are small.}

This model offers no natural explanation of the steady dimming
reported by \citet{MS16}.  It might be explained as a consequence of a
broader ring of dilute obscuring material, but that adds an {\it ad hoc\/}
complexity to the model. 
\section*{Acknowledgements}
I thank {D. Palmer, S. Phinney and} S. A. Rappaport for useful
discussions { on every aspect of this problem and an anonymous referee
for much constructive criticism.  This research has made use of the NASA
Exoplanet Archive, which is operated by the California Institute of
Technology, under contract with the National Aeronautics and Space
Administration under the Exoplanet Exploration Program}.

\bsp 
\label{lastpage} 
\end{document}